\documentclass[12pt,english,draftcls, one column]{IEEEtran}

\usepackage[T1]{fontenc}
\usepackage[latin9]{inputenc}
\usepackage{float}
\usepackage{amsmath}
\usepackage{amsthm}
\usepackage{amssymb}
\usepackage{graphicx}
\usepackage{stfloats}
\usepackage{cite} 
\usepackage{xcolor}
\usepackage{bbm}
\usepackage{adjustbox}

\makeatletter

\floatstyle{ruled}
\newfloat{algorithm}{tbp}{loa}
\providecommand{\algorithmname}{Algorithm}
\floatname{algorithm}{\protect\algorithmname}

\theoremstyle{plain}

\theoremstyle{plain}

\IEEEoverridecommandlockouts
\usepackage{cite}
\usepackage{amsfonts}\usepackage{algorithmic}
\usepackage{textcomp}
\usepackage{etoolbox}

\def\BibTeX{{\rm B\kern-.05em{\sc i\kern-.025em b}\kern-.08em
    T\kern-.1667em\lower.7ex\hbox{E}\kern-.125emX}}

\patchcmd{\@maketitle}
  {\addvspace{0.5\baselineskip}\egroup}
  {\addvspace{-1\baselineskip}\egroup}
  {}
  {}

\makeatother

\providecommand{\propositionname}{Proposition}
\providecommand{\theoremname}{Theorem}

\begin{document}

\title{\huge Generalized Reduced-WMMSE Approach for Cell-Free Massive MIMO With\\ Per-AP Power Constraints}

\author{Wonsik Yoo, \textit{Graduate Student Member}, \textit{IEEE}, Daesung Yu, \textit{Member}, \textit{IEEE}, Hoon Lee, \textit{Member}, \textit{IEEE}, and Seok-Hwan Park, \textit{Senior Member}, \textit{IEEE} \thanks{
This work is supported in part by the National Research Foundation (NRF) of Korea Grants funded by the MOE under Grant 2019R1A6A1A09031717 and 2021R1I1A3054575 and by the MSIT under Grant RS-2023-00238977. \textit{(Corresponding authors: Hoon Lee and Seok-Hwan Park.)}


W. Yoo and S.-H. Park are with the Division of Electronic Engineering, Jeonbuk
National University, Jeonju, Korea (email: wonsik0713@jbnu.ac.kr, seokhwan@jbnu.ac.kr).

D. Yu is with the Institute for Communication Systems, University of Surrey, Guildford, UK (email: d.yu@surrey.ac.uk).

H. Lee is with the Department of Electrical Engineering and Artificial Intelligence Graduate School, Ulsan National Institute of Science and Technology (UNIST), Ulsan, Korea (email: hoonlee@unist.ac.kr).
}
}
\maketitle
\begin{abstract}
The optimization of cooperative beamforming vectors in cell-free massive MIMO (mMIMO) systems is presented where multi-antenna access points (APs) support downlink data transmission of multiple users. Albeit the successes of the weighted minimum mean squared error (WMMSE) algorithm and their variants, they lack careful investigations about computational complexity that scales with the number of antennas and APs. We propose a generalized and reduced WMMSE (G-R-WMMSE) approach whose complexity is significantly lower than conventional WMMSE.
We partition the set of beamforming coefficients into subvectors, with each subvector corresponding to a specific AP. Such a partitioning approach decomposes the original WMMSE problem across individual APs. By leveraging the Lagrange duality analysis, a closed-form solution can be derived for each subproblem, which substantially reduces the computation burden.
Additionally, we present a parallel execution of the proposed G-R-WMMSE with adaptive step sizes, aiming at further reducing the time complexity.
Numerical results validate that the proposed G-R-WMMSE schemes achieve over 99$\%$ complexity savings compared to the conventional WMMSE scheme while maintaining almost the same performance.
\end{abstract}


\theoremstyle{theorem}
\newtheorem{theorem}{Theorem} 
\theoremstyle{proposition}
\newtheorem{proposition}{Proposition} 
\theoremstyle{lemma}
\newtheorem{lemma}{Lemma} 
\theoremstyle{corollary}
\newtheorem{corollary}{Corollary} 
\theoremstyle{definition}
\newtheorem{definition}{Definition}
\theoremstyle{remark}
\newtheorem{remark}{Remark}

\section{Introduction}

Cell-free massive MIMO (mMIMO) system is envisioned to play a pivotal role in attaining superior spectral and energy efficiency in the next-generation wireless systems \cite{Ngo:TWC17, Nayebi:TWC17, Bjornson:TCOM20, Mai:TCOM20}.
To effectively manage interference signals of cell-free mMIMO systems, cooperative beamforming across distributed access points (APs) controlled by a central processor (CP) is imperative.
Among various beamforming optimization methods, the weighted minimum mean squared error (WMMSE) algorithm has gained popularity as it provides effective beamforming vectors across a range of criteria including sum-rate maximization \cite{Christensen:TWC08}.
Nonetheless, the high computational complexity associated with the WMMSE algorithm renders it less suitable for cell-free mMIMO systems involving a multitude of APs.
In addressing this issue, an efficient alternative referred to as the reduced-WMMSE (R-WMMSE) algorithm was introduced in \cite{Zhao:TSP23}.
This approach finds locally optimal solutions for weighted sum-rate maximization problems in conventional single-cell networks where a single AP serves several users. Both the sum-power constraint (SPC) and per-antenna power constraint (PAPC) were considered. For these cases, the R-WMMSE method has been shown to exhibit significantly lower computational complexity compared to the original WMMSE scheme without any performance loss.

Inspired by these results, this paper proposes the R-WMMSE algorithm for solving cooperative beamforming problems in the cell-free mMIMO networks. Unlike the single-cell system \cite{Zhao:TSP23}, the cell-free mMIMO imposes per-AP power constraints that restrict the transmit power of a group of antennas deployed to individual APs. For this reason, the SPC and PAPC investigated in \cite{Zhao:TSP23} are respectively regarded as special cases of the cell-free mMIMO systems with a single multi-antenna AP, i.e., all transmit antennas are co-located, and multiple single-antenna APs, i.e., each of antennas are fully separated. To cope with a generic multi-antenna multi-AP network, it is essential to develop a new R-WMMSE algorithm which can generalize the SPC and PAPC scenarios. 
The WMMSE method was also adopted in the cell-free mMIMO for optimizing power control variables with fixed beamforming \cite{Chakraborty:OJCS21} and under the fronthaul capacity constraint~\cite{Xu:OJCS22}.

To address these challenges, we develop a generalized R-WMMSE (G-R-WMMSE) scheme.
In the cell-free mMIMO systems, the beamforming vectors are designed under per-AP power constraints \cite{Zhang:JSAC10}, which are distinct constraints not covered by the SPC or PAPC addressed in \cite{Zhao:TSP23}.
We partition the set of beamforming coefficients into subvectors, with each subvector corresponding to a specific AP. As a result, the WMMSE problem can be decoupled into multiple subproblems, where each subproblem is dedicated to optimizing the beamforming vector at a certain AP.
We then employ the block coordinate descent (BCD) approach \cite[Sec. 2.7]{Bertsekas:Cambridge99} to iteratively optimize these partitioned subvectors via an alternating optimization procedure of subproblems.
Recognizing that the subproblem is convex with the strong duality, we leverage the Lagrange duality method \cite{Boyd:Cambridge04} to obtain a closed-form solution, contributing to the complexity reduction.
Additionally, we present a parallel execution strategy of the proposed G-R-WMMSE. Unlike the normal BCD algorithm that sequentially solves subproblems, this method allows simultaneous updates of individual beamforming vectors, thereby reducing the time complexity by means of parallel computation.
Numerical results confirm that the G-R-WMMSE schemes offer performance on par with the conventional WMMSE method with reduced complexity.

\section{System Model} \label{sec:System-Model}

We consider the downlink of a cell-free mMIMO system comprising $K$ single-antenna user equipments (UEs), $M$ APs, and a CP.
Each AP is equipped with $n_A$ antennas.
The communication between UEs and APs occurs over a wireless access link, while each AP $i$ is connected to the CP through a fronthaul link.
In this work, we assume ideal fronthaul links which do not cause any distortion, and the design with limited fronthaul is left for future work.
We use the sets $\mathcal{K} = \{1,2,\ldots,K\}$ and $\mathcal{M} = \{1,2,\ldots,M\}$ to denote the collections of the UEs' and APs' indices, respectively.

\subsection{Uplink Channel Training and Imperfect CSI Model} \label{sub:uplink-training}

We employ a time-division duplexing (TDD) protocol, where the acquisition of the downlink channel state information (CSI) relies on an uplink channel training process, making use of channel reciprocity \cite{Ngo:TWC17, Nayebi:TWC17, Bjornson:TCOM20, Mai:TCOM20}.
The channel vector between UE $k$ and AP $i$ is denoted by $\mathbf{h}_{k,i}\sim \mathcal{CN}(\mathbf{0}, \rho_{k,i}\mathbf{I})\in\mathbb{C}^{n_A\times 1}$, where $\rho_{k,i}$ is the large-scale fading coefficient. There are $L$ orthogonal pilot sequences each having length $L$. UE $k$ chooses the $l_{k}$th pilot signal ($l_{k}=1,\cdots,L$) and transmits it to the APs. Without loss of the generality, it is assumed that several UEs can select the identical pilot, thereby leading to the pilot contamination issue. The set of the UEs transmitting $l$th pilot is denoted by $\mathcal{K}_{l}\subset\mathcal{K}$.

Using the received pilot signals, each AP $i$ obtains an estimate $\{\hat{\mathbf{h}}_{k,i}\}_{k\in\mathcal{K}}$ of its local CSI $\{\mathbf{h}_{k,i}\}_{k\in\mathcal{K}}$.
With the linear MMSE estimator, $\hat{\mathbf{h}}_{k,i}$ is modeled as
\begin{align}
    \mathbf{h}_{k,i} = \hat{\mathbf{h}}_{k,i} + \tilde{\mathbf{h}}_{k,i}, \label{eq:CSI-error-model}
\end{align}
where the estimation error $\tilde{\mathbf{h}}_{k,i}\sim\mathcal{CN}(\mathbf{0}, \tilde{\rho}_{k,i}\mathbf{I})$ is uncorrelated with $\hat{\mathbf{h}}_{k,i}$.
The error variance $\tilde{\rho}_{k,i}$ is given by $\tilde{\rho}_{k,i} = \rho_{k,i} - \hat{\rho}_{k,i}$, where $\hat{\rho}_{k,i}$ denotes the variance of the estimated channel $\hat{\mathbf{h}}_{k,i}$ obtained as \cite{Shaik:TCOM21}
\begin{align}
    \hat{\rho}_{k,i} = \frac{ \rho_{k,i}^2 }{ 1/\left(L\frac{P^{\text{ul}}}{N_0^{\text{ul}}}
    \right) + \sum_{l\in\mathcal{K}_{l_k}} \rho_{l,i}}. \label{eq:nominal-CSI-variance}
\end{align}
Here, $P^{\text{ul}}$ and $N_0^{\text{ul}}$ equal the uplink pilot transmission power and the noise variance at the APs, respectively.

\subsection{Downlink Payload Data Transmission} \label{sub:downlink-payload}

In the downlink payload phase, the CP generates channel-encoded baseband signal $s_k\sim\mathcal{CN}(0, 1)$ for each UE $k\in\mathcal{K}$.
These data signals are linearly precoded for interference management. As a result, the transmitted signal vector $\mathbf{x}\in\mathbb{C}^{M n_A \times 1}$ across all APs is expressed as
\begin{align}
    \mathbf{x} = \sum\nolimits_{k\in\mathcal{K}} \mathbf{v}_k s_k, \label{eq:beamforming}
\end{align}
where $\mathbf{v}_k = [\mathbf{v}_{k,1}^H \, \cdots \, \mathbf{v}_{k,M}^H]^H\in\mathbb{C}^{M n_A\times 1}$ represents the beamforming vector for $s_k$ with the subvector $\mathbf{v}_{k,i}\in\mathbb{C}^{n_A\times 1}$ corresponding to AP $i$. 
Denoting  the $i$th subvector of $\mathbf{x}$ transmitted by AP $i$ by $\mathbf{x}_i\in\mathbb{C}^{n_A\times 1}$, we impose per-AP transmit power constraints as
\begin{align}
    \mathbb{E}[\|\mathbf{x}_i\|^2] = \sum\nolimits_{k\in\mathcal{K}} \|\mathbf{v}_{k,i}\|^2 \leq P^{\text{dl}}_i, \, i\in\mathcal{M}, \label{eq:power-constraint}
\end{align}
where $P^{\text{dl}}_i$ is the downlink power budget of AP $i$.

The received signal $y_k$ of UE $k$ can be expressed as
\begin{align}
    y_k \!=\! \hat{\mathbf{h}}_k^H \mathbf{v}_k s_k \!+\! \sum\nolimits_{l\in\mathcal{K}\setminus\{k\}} \!\hat{\mathbf{h}}_k^H\mathbf{v}_l s_l \!+\! \sum\nolimits_{l\in\mathcal{K}}\! \tilde{\mathbf{h}}_k^H \mathbf{v}_l s_l \!+\! z_k, \label{eq:received-signal-downlink}
\end{align}
where $\hat{\mathbf{h}}_k = [\hat{\mathbf{h}}_{k,1}^H \, \cdots \, \hat{\mathbf{h}}_{k,M}^H]^H$ and $\tilde{\mathbf{h}}_k = [\tilde{\mathbf{h}}_{k,1}^H \, \cdots \, \tilde{\mathbf{h}}_{k,M}^H]^H$ respectively stack the estimated CSI and error vectors and $z_k\sim\mathcal{CN}(0, N_0^{\text{dl}})$ is the additive noise with variance $N_0^{\text{dl}}$.
The third term represents the interference caused by channel estimation error.

Given the uncertainty in the channel estimation error $\tilde{\mathbf{h}}_k$, we consider the expected data rate averaged over the estimation error vector $\tilde{\mathbf{h}}_k$. According to \cite{Pan:TWC18}, the closed-form approximation for the expected rate of UE $k$ can be derived~as
\begin{align}
    &R_k = f_k\left( \mathbf{v} \right) = \log_2\left( 1 +  \frac{ \big| \hat{\mathbf{h}}_k^H\mathbf{v}_k \big|^2 }{ \text{IF}_k\left(\mathbf{v}\right)+N_{0}^{\text{dl}}} \right), \label{eq:data-rate-lower-bound}
\end{align}
where $\mathbf{v} = \{\mathbf{v}\}_{k\in\mathcal{K}}$ collects all beamforming vectors and the interference power $\text{IF}_k(\mathbf{v})$ is defined as
\begin{align}
    \text{IF}_k(\mathbf{v}) =  \sum\nolimits_{l\in\mathcal{K}\setminus\{k\}} |\hat{\mathbf{h}}_k^H\mathbf{v}_l |^2 + \sum\nolimits_{l\in\mathcal{K}} \mathbf{v}_l^H\mathbf{E}_k\mathbf{v}_l
\end{align}
with $\mathbf{E}_k = \text{blkdiag}(\{\tilde{\rho}_{k,i}\mathbf{I}_{n_A}\}_{i\in\mathcal{M}})$ being the CSI error covariance. Note that the closed-form expression $R_k$ presented above serves as a lower bound on the expected rate.

In this work, we tackle the problem of maximizing the weighted sum-rate objective function $\sum_{k\in\mathcal{K}} \mu_k f_{k}(\mathbf{v})$ by optimizing the beamforming vectors $\mathbf{v}$, where $\mu_k > 0$ represents the priority of the quality-of-service (QoS) for UE $k$. 
We can formulate the problem as
\begin{subequations} \label{eq:problem-original}
\begin{align}
    \underset{\mathbf{v}} {\mathrm{max.}}\,\,\, & \sum\nolimits_{k\in\mathcal{K}} \mu_k f_k\left( \mathbf{v} \right) \, \label{eq:problem-original-objective} \\
 \mathrm{s.t. }\,\,\,\,\,\, & \sum\nolimits_{k\in\mathcal{K}} \|\mathbf{v}_{k,i}\|^2  \leq P^{\text{dl}}_i, \, i\in\mathcal{M}. \label{eq:problem-original-power}
\end{align}
\end{subequations}

\section{Conventional WMMSE Approach} \label{sec:conventional-WMMSE}

This section presents the conventional WMMSE method to solve \eqref{eq:problem-original}. We define the mean squared error (MSE) between the signal $s_k$ and $u_k^Hy_k$ with a scalar receive filter $u_k\in\mathbb{C}$ as

\begin{align}
\text{MSE}_k\left( \mathbf{v}, u_k \right) &= \mathbb{E}\left[ 
|s_k - u_k^H y_k|^2 \right] \label{eq:definition-MSE} \\
&= \big|1 - u_k^H\hat{\mathbf{h}}_k^H\mathbf{v}_k\big|^2 + |u_k|^2\left(\text{IF}_k\left(\mathbf{v}\right) + N_{0}^{\text{dl}}\right). \nonumber
\end{align}

For arbitrary receive filter $u_k$ and positive weight $w_k > 0$, we have the following inequality \cite{Christensen:TWC08}.
\begin{align}
    f_k\left(\mathbf{v}\right) \geq \log_2 w_k -\frac{w_k}{\ln 2} \text{MSE}_k\left( \mathbf{v}, u_k \right) + \frac{1}{\ln 2}, \label{eq:WMMSE-inequality}
\end{align}
where the equality holds when
\begin{align}
    \!\!u_k = \frac{\hat{\mathbf{h}}_k^H\mathbf{v}_k}{ \big|\hat{\mathbf{h}}_k^H\mathbf{v}_k\big|^2 + \text{IF}_k\left(\mathbf{v}\right) + N_{0}^{\text{dl}}} \,\,\,\text{and} \,\,\,
    w_k = \frac{1}{1 - u_k^H\hat{\mathbf{h}}_k^H\mathbf{v}_k}. 
    \label{eq:optimal-u-w}
\end{align}

Using the inequality (\ref{eq:WMMSE-inequality}), we can recast (\ref{eq:problem-original}) to
\begin{subequations} \label{eq:problem-WMMSE}
\begin{align}
    \underset{\mathbf{v}, \mathbf{u},\mathbf{w}} {\mathrm{min.}}\,\,\, & \sum\nolimits_{k\in\mathcal{K}} \mu_k \left( \frac{w_k}{\ln 2} \text{MSE}_k\left( \mathbf{v}, u_k \right) - \log_2 w_k \!\right) \, \label{eq:problem-WMMSE-objective} \\
 \mathrm{s.t. }\,\,\,\,\,\, & \sum\nolimits_{k\in\mathcal{K}} \|\mathbf{v}_{k,i}\|^2 \leq P^{\text{dl}}_i, \, i\in\mathcal{M}, \label{eq:problem-WMMSE-power}
\end{align}
\end{subequations}
with $\mathbf{u} = \{u_k\}_{k\in\mathcal{K}}$ and $\mathbf{w} = \{w_k\}_{k\in\mathcal{K}}$.
For fixed $\{\mathbf{u},\mathbf{w}\}$, problem (\ref{eq:problem-WMMSE}) with respect to $\mathbf{v}$ is convex. Furthermore, the optimal values of $\{\mathbf{u},\mathbf{w}\}$, that minimize the cost function for fixed $\mathbf{v}$, can be obtained in closed-form as shown in (\ref{eq:optimal-u-w}).
An alternating optimization procedure between $\mathbf{v}$ and $\{\mathbf{u},\mathbf{w}\}$, termed by the WMMSE algorithm, monotonically improves the sum-rate performance in iterations. The WMMSE method has been shown to find a locally optimal solution to (\ref{eq:problem-original}).
As will be discussed in Sec. \ref{sec:complexity}, the complexity of the conventional WMMSE algorithm becomes notably burdensome, especially when $K$, $M$, and $n_A$ grows larger.

\section{Proposed G-R-WMMSE Approach} \label{sec:proposed}

To address the complexity challenge, we present an efficient WMMSE algorithm whose complexity is significantly lower than the conventional WMMSE scheme. A key idea is to extend the R-WMMSE method \cite{Zhao:TSP23} originally intended for a single-AP system under the SPC or the PAPC. These two extreme power constraints are viewed as special cases of the per-AP power constraint \eqref{eq:power-constraint} which invokes generalized power constraints for a group of antennas. If all the transmit antennas are co-located at a single AP, i.e., $M=1$ and $n_{A}=Mn_{A}$, problem \eqref{eq:problem-original} boils down to the beamforming optimization of a single multi-antenna AP $M=1$ with the SPC. Also, the PAPC setting is obtained with single-antenna APs $n_{A}=1$. Therefore, the cell-free mMIMO problem in \eqref{eq:problem-original} requests a generalization of the conventional R-WMMSE algorithm \cite{Zhao:TSP23}.

\subsection{Sequential G-R-WMMSE} \label{sub:sequential-G-R-WMMSE}
The proposed G-R-WMMSE algorithm addresses the WMMSE minimization task in \eqref{eq:problem-WMMSE} for fixed $\{\mathbf{u},\mathbf{w}\}$. We first partition $\mathbf{v}$ into $M$ blocks of variables $\mathbf{v}_{A,i}\triangleq[\mathbf{v}_{1,i}^H \, \cdots \, \mathbf{v}_{K,i}^H]^H \in \mathbb{C}^{n_A K \times 1}$, $\forall i\in\mathcal{M}$, where each block $\mathbf{v}_{A,i}$ collects beam weights associated with AP $i$. The BCD algorithm \cite[Sec. 2.7]{Bertsekas:Cambridge99} is employed to alternately identify each block $\mathbf{v}_{M,i}$ by fixing others $\mathbf{v}_{A,j}$, $\forall j\in\mathcal{M}\setminus\{i\}$. Then, we can split the problem in \eqref{eq:problem-WMMSE} into $M$ subproblems each of which is dedicated to the optimization of the beamforming vector $\mathbf{v}_{A,i}$ of AP $i$. The $i$th subproblem is obtained as
\begin{subequations} \label{eq:problem-AP-i}
\begin{align}
    \underset{\mathbf{v}_{A,i}} {\mathrm{min.}}\,\,\, & \mathbf{v}_{A,i}^H \mathbf{Q}_i \mathbf{v}_{A,i} + 2 \Re\big\{\mathbf{b}_i^H\mathbf{v}_{A,i}\big\} \, \label{eq:problem-AP-i-objective} \\
 \mathrm{s.t. }\,\,\,\, & \|\mathbf{v}_{A,i}\|^2 \leq P_i^{\text{dl}}, \label{eq:problem-AP-i-power}
\end{align}
\end{subequations}
where we have defined $\mathbf{Q}_i \succ \mathbf{0}$ and $\mathbf{b}_i$ as
\begin{subequations} \label{eq:Q-i-b-i}
\begin{align}
    &\mathbf{Q}_i =  \mathbf{I}_K \otimes \left( \hat{\mathbf{H}}_i\mathbf{A}\hat{\mathbf{H}}_i^H \right) + \mathbf{C}_i,\label{eq:Q-i} \\
    &\mathbf{b}_i \!=\! -\mathbf{B} \hat{\mathbf{h}}_{A,i} \!+\! \sum\nolimits_{m\in\mathcal{M}\setminus\{i\}} \!\! \left( \mathbf{I}_K \!\otimes\! \left( \hat{\mathbf{H}}_i\mathbf{A}\hat{\mathbf{H}}_m^H \! \right) \right) \! \mathbf{v}_{A,m}, \label{eq:b-i}
\end{align}
\end{subequations}
with $\hat{\mathbf{H}}_m = [\hat{\mathbf{h}}_{1,m} \, \cdots \,\hat{\mathbf{h}}_{K,m}]$, $\hat{\mathbf{h}}_{A,i} = [\hat{\mathbf{h}}_{1,i}^H \, \cdots \, \hat{\mathbf{h}}_{K,i}^H]^H$, and
\begin{subequations}
\begin{align}
    \mathbf{A} &= \text{diag} (\{ \mu_k w_k|u_k|^2 \}_{k\in\mathcal{K}}),\\
    \mathbf{B} &= \text{blkdiag}(\{\mu_k w_k u_k{\mathbf{I}_{n_A}}\}_{k\in\mathcal{K}}), \\
    \mathbf{C}_i &= \Big( \sum\nolimits_{k\in\mathcal{K}} \mu_k w_k|u_k|^2 \tilde{\rho}_{k,i} \Big)\mathbf{I}_{n_A K}.
\end{align}
\end{subequations}

Thanks to the convexity and the strong duality, \eqref{eq:problem-AP-i} can be optimally addressed by using the Lagrange duality method \cite{Boyd:Cambridge04}.
The Lagrangian of (\ref{eq:problem-AP-i}) is written as
\begin{align}
    \mathcal{L}\left(\mathbf{v}_{A,i}, \lambda_i\right) = \mathbf{v}_{A,i}^H \mathbf{Q}_i \mathbf{v}_{A,i} + 2 \Re\big\{\mathbf{b}_i^H\mathbf{v}_{A,i}\big\} + \lambda_i\left( \|\mathbf{v}_{A,i}\|^2 - P_i^{\text{dl}} \right), \label{eq:Lagrangian}
\end{align}
with a Lagrange multiplier $\lambda_i \geq 0$.
By imposing the first-order KKT conditions, the beamforming solution $\mathbf{v}_{A,i}$ for minimizing the Lagrangian is given as 
\begin{align}
    \mathbf{v}_{A,i} = -\left(\mathbf{Q}_i + \lambda_i \mathbf{I}_{n_A K}\right)^{-1}\mathbf{b}_i. \label{eq:optimal-v-tilde-i}
\end{align}
Plugging \eqref{eq:optimal-v-tilde-i} to \eqref{eq:Lagrangian} leads to the dual function obtained as
\begin{align}
    g(\lambda_{i})=-\mathbf{b}_{i}^{H}(\mathbf{Q}_{i}+\lambda_{i}\mathbf{I}_{n_{A}K})^{-1}\mathbf{b}_{i}-P_{i}^{\text{dl}}\lambda_{i}.
\end{align}
Then, the dual problem is formulated as $\max_{\lambda_{i}\geq0} g(\lambda_{i})$.
The zero-gradient condition $\partial g(\lambda_{i})/\partial \lambda_{i}=0$ reveals that the optimal dual variable $\lambda_{i}$ satisfies the following condition:
\begin{align}
    p_{i}(\lambda_{i})\triangleq\mathbf{b}_{i}^{H}(\mathbf{Q}_{i}+\lambda_{i}\mathbf{I}_{n_{A}K})^{-2}\mathbf{b}_{i}=P_{i}^{\text{dl}}. \label{eq:optimal-Lagrange-multiplier}
\end{align}
Since $p_i(\lambda_i)$ is monotonically decreasing, when $p_{i}(0)<P_{i}^{\text{dl}}$, \eqref{eq:optimal-Lagrange-multiplier} results in a negative dual variable, which is infeasible for the KKT condition $\lambda_{i}\geq0$. Thus, in this case, the optimal $\lambda_{i}$ becomes $\lambda_{i}=0$. Otherwise, the dual variable fulfilling \eqref{eq:optimal-Lagrange-multiplier} can be found by using the bisection method \cite[Sec. 4.2.5]{Boyd:Cambridge04}.

\begin{remark}
    If each AP has a single antenna, i.e., $n_A=1$, the matrix $\mathbf{Q}_i$ in (\ref{eq:Q-i}) reduces to $\mathbf{Q}_i = \alpha_i \mathbf{I}_K$ where $\alpha_{i} = \sum_{k\in\mathcal{K}}w_k|u_k|^2 (|\hat{h}_{k,i}|^2 + \tilde{\rho}_{k,i})$ with $\hat{h}_{k,i}$ being a scalar version of $\hat{\mathbf{h}}_{k,i}$ when $n_A = 1$. Thus, (\ref{eq:problem-AP-i}) has a scaled-norm property leading to a closed-form solution under the PAPC \cite{Zhao:TSP23}:
    \begin{align}
        \mathbf{v}_{A,i} = -\mathbf{b}_i \min \left( \frac{1}{\alpha_i}, \frac{\sqrt{P_i^{\text{dl}}}}{\|\mathbf{b}_i\|} \right). \label{eq:closed-form-solution-Zhao}
    \end{align}
    Since a bisection search is not required in (\ref{eq:closed-form-solution-Zhao}), it is simpler than our solution (\ref{eq:optimal-v-tilde-i}) designed for arbitrary number of AP antennas $n_A$. Thus, the proposed solution generalizes (\ref{eq:closed-form-solution-Zhao}) to the scenario with per-AP power constraints.
\end{remark}

The complete process of the proposed G-R-WMMSE method is outlined in Algorithm 1. A key difference from the conventional WMMSE \cite{Christensen:TWC08} is the beamforming update steps in lines 6-7, in which the beamforming vectors are alternately optimized on a per-AP basis instead of being jointly optimized.
As the block variables $\mathbf{v}_{A,i}$, $i\in\mathcal{M}$, are sequentially updated, we refer to Algorithm 1 as \textit{Sequential G-R-WMMSE}.

\begin{algorithm}
\caption{Sequential G-R-WMMSE algorithm}

\textbf{\footnotesize{}1}\textbf{ Initialize}

\textbf{\footnotesize{}2}~~~~Set $\mathbf{v}$ to arbitrary beamforming vectors that satisfy (\ref{eq:problem-original-power}) and store $\mathbf{v}^{\text{old}}\leftarrow \mathbf{v}$.

\textbf{\footnotesize{}3}~~~~Update $\{\mathbf{u},\mathbf{w}\}$ as (\ref{eq:optimal-u-w}).

\textbf{\footnotesize{}4}\textbf{ Repeat}

\textbf{\footnotesize{}5}~~~~For $i \leftarrow 1$ to $M$ (\textbf{sequentially})

\textbf{\footnotesize{}6}~~~~~~Calculate $\mathbf{Q}_i$ and $\mathbf{b}_i$ with (\ref{eq:Q-i-b-i}).

\textbf{\footnotesize{}7}~~~~~~Update $\mathbf{v}_{A,i}$ with (\ref{eq:optimal-v-tilde-i}).

\textbf{\footnotesize{}8}~~~~End

\textbf{\footnotesize{}9}~~~~Update $\{\mathbf{u},\mathbf{w}\}$ as (\ref{eq:optimal-u-w}).

\textbf{\footnotesize{}10} \textbf{Until} $\sum_{i\in\mathcal{M}}\| \mathbf{v}_{A,i} - \mathbf{v}_{A,i}^{\text{old}} \|^2_F \leq \delta$ (Otherwise, set $\mathbf{v}^{\text{old}} \leftarrow \mathbf{v}$)

\end{algorithm}

\subsection{Parallel G-R-WMMSE} \label{sub:parallel-G-R-WMMSE}

In this subsection, we propose a parallel version of the proposed G-R-WMMSE algorithm, referred to as \textit{Parallel G-R-WMMSE}, that adopts
parallel update of block variables $\mathbf{v}_{A,i}$, $i\in\mathcal{M}$, with the aim of further speeding up the algorithm execution.
To this end, we first rewrite $\mathbf{b}_i$ in (\ref{eq:b-i}) as
\begin{align}
    \mathbf{b}_i\! =\! -\mathbf{B} \hat{\mathbf{h}}_{A,i} + (\mathbf{I}_K\otimes \hat{\mathbf{H}}_i)(\mathbf{d} - ( 
\mathbf{I}_K \otimes (\mathbf{A} \hat{\mathbf{H}}_i^H) )\mathbf{v}_{A,i}),\label{eq:b_i_new}
\end{align}
where the vector $\mathbf{d}\in\mathbb{C}^{K^2\times1}$ is defined as
\begin{align}
    \mathbf{d} =\sum\nolimits_{m\in\mathcal{M}}\text{vec}(\mathbf{A}\hat{\mathbf{H}}_{m}^{H}\mathbf{V}_{m}) \label{eq:D}
\end{align}
with $\mathbf{V}_{m}\triangleq[\mathbf{v}_{1,m}\cdots\mathbf{v}_{K,m}]\in\mathbb{C}^{n_{A}\times K}$ and $\text{vec}(\cdot)$ being the vectorization operator.
Thus, once $\mathbf{d}$ is computed with a linear complexity in the number of APs $M$, each vector $\mathbf{b}_i$ can be obtained via (\ref{eq:b_i_new}) whose complexity is not affected by $M$.

However, if we simultaneously update $\mathbf{v}_{A,i}$, $i\in\mathcal{M}$, with (\ref{eq:optimal-v-tilde-i}), it might result in severe fluctuation of sum-rate with respect to iterations. This is due to the fact that the vector $\mathbf{b}_i$ of AP $i$ depends on beamforming vectors of others $\{\mathbf{v}_j\}_{j\in\mathcal{M}\setminus\{i\}}$. Parallel updates induce huge variations in these beamforming vectors since their updated versions can only be available in the subsequent iteration.
To address this issue, we adopt the approach of the exact Jacobi best-response scheme \cite[Sec. IV]{Scutari:TSP14}. The beamforming vector $\mathbf{v}_{A,i}$ is gradually updated as 
\begin{align}
\mathbf{v}_{A,i} \leftarrow \beta \mathbf{v}_{A,i}^{\text{new}} + (1-\beta)\mathbf{v}_{A,i}^{\text{old}},
\end{align}
where $\mathbf{v}_{A,i}^{\text{new}}$ corresponds to the best-response solution obtained by solving (\ref{eq:optimal-v-tilde-i}) at each iteration, $\mathbf{v}_{A,i}^{\text{old}}$ stands for the previous solution, and $\beta\in[0,1]$ is the step size controlling the speed of updates.
Notice that by setting $\beta=1$, this boils down to the beamforming update rule $\mathbf{v}_{A,i} \leftarrow \mathbf{v}_{A,i}^{\text{new}}$ of the sequential G-R-WMMSE algorithm.
At the end of each iteration, we adjust the step size as $\beta\leftarrow\beta(1-\epsilon\beta)$ with a given $\epsilon\in[0,1]$.
This allows for gradual reduction of the step size over iterations.
The value of $\epsilon$, which controls the pace of step size reduction, is chosen while balancing between convergence speed and algorithmic stability, as discussed in \cite[Sec. IV]{Scutari:TSP14}.
We summarize the parallel G-R-WMMSE method in Algorithm~2. 

\begin{algorithm}
\caption{Parallel G-R-WMMSE algorithm}

\textbf{\footnotesize{}1}\textbf{ Initialize}

\textbf{\footnotesize{}2}~~~~Set $\mathbf{v}$ to arbitrary beamforming vectors that satisfy (\ref{eq:problem-original-power})

~~~~~and store $\mathbf{v}^{\text{old}}\leftarrow \mathbf{v}$.

\textbf{\footnotesize{}3}~~~~Update $\{\mathbf{u},\mathbf{w}\}$ as (\ref{eq:optimal-u-w}).

\textbf{\footnotesize{}4}\textbf{ Repeat}

\textbf{\footnotesize{}5}~~~~Calculate $\mathbf{d}$ with (\ref{eq:D}).

\textbf{\footnotesize{}6}~~~~For $i \in \mathcal{M}$ (\textbf{in parallel})

\textbf{\footnotesize{}7}~~~~~~Calculate $\mathbf{Q}_i$ and $\mathbf{b}_i$ with (\ref{eq:Q-i}) and (\ref{eq:b_i_new}).

\textbf{\footnotesize{}8}~~~~~~Calculate $\mathbf{v}_{A,i}^{\text{new}}$ with (\ref{eq:optimal-v-tilde-i}).

\textbf{\footnotesize{}9}~~~~~~Update $\mathbf{v}_{A,i} \leftarrow \beta \mathbf{v}_{A,i}^{\text{new}} + (1-\beta)\mathbf{v}_{A,i}^{\text{old}}$.

\textbf{\footnotesize{}10}~~~~End

\textbf{\footnotesize{}11}~~~~Update $\{\mathbf{u},\mathbf{w}\}$ as (\ref{eq:optimal-u-w}).

\textbf{\footnotesize{}12} \textbf{Until} $\sum_{i\in\mathcal{M}}\| \mathbf{v}_{A,i} - \mathbf{v}_{A,i}^{\text{old}} \|^2_F \leq \delta$ 

\textbf{\footnotesize{}13}
~~~~~~~(Otherwise, set $\mathbf{v}^{\text{old}} \leftarrow \mathbf{v}$ and $\beta\leftarrow\beta(1-\epsilon\beta)$)

\end{algorithm}

\subsection{Complexity Discussion} \label{sec:complexity}

The complexity of conventional WMMSE at each iteration is dominated by that of solving the convex problem (\ref{eq:problem-WMMSE}) for fixed $\{\mathbf{u}, \mathbf{w}\}$, which is given by $\mathcal{O}(K^4 M^4 n_A^4)$. This is obtained from $\mathcal{O}(n_V (n_V^3 + n_O))$ \cite[p. 4]{Ben-Tal}, where $n_V = \mathcal{O}(K M n_A)$ and $n_O = \mathcal{O}(K^2 n_A M)$ respectively denote the numbers of optimization variables and the arithmetic operations required to compute the objective and constraint functions.
On the contrary, the proposed G-R-WMMSE algorithms execute closed-form computations, and the complexity becomes $\mathcal{O}(K^3 (M^2 n_A + M n_A^3))$ and $\mathcal{O}(K^3 (M n_A + n_A^3))$ for the sequential and parallel implementations, respectively.
They have notably lower growth with respect to $K$, $M$, and $n_A$ compared to conventional WMMSE. Furthermore, for given $K$ and $n_A$, the parallel G-R-WMMSE reduces the complexity of the sequential G-R-WMMSE scheme from $\mathcal{O}(M^2)$ to $\mathcal{O}(M)$, offering a significant computational advantage.

\section{Numerical Results} \label{sec:numerical-results}

We present numerical results to validate the advantages of the proposed G-R-WMMSE scheme.
The UEs and APs are uniformly distributed within a circular area of radius 350 m. 
The path-loss $\rho_{i,k}$ is modeled as $\rho_{i,k} = (d_{i,k}/d_0)^{-\eta}$ with the reference distance $d_0 = 30$ m and path-loss exponent $\eta = 3$, where $d_{i,k}$ is the distance between UE $k$ and AP $i$. All the APs have the same transmit power budget, i.e., $P_i^{\text{dl}} = P^{\text{dl}}$, $\forall i\in\mathcal{M}$.
Orthogonal pilot sequences are assigned to the UEs in a round-robin fashion, i.e., $l_k = \left((k-1) \mod L\right) + 1$ for $k\in\mathcal{K}$. The downlink and uplink SNRs are respectively defined as $\text{SNR}^{\text{dl}}\triangleq P^{\text{dl}}/N_0^{\text{dl}}$ and $\text{SNR}^{\text{ul}}\triangleq P^{\text{ul}}/N_0^{\text{ul}}$.
We set $\epsilon = 0.1$ in the parallel G-R-WMMSE scheme.
Unless stated otherwise, we consider the sum-rate metric with $\mu_k=1$, $\forall k\in\mathcal{K}$.

\begin{figure}
\centering\!\!\!\!\!\includegraphics[width=0.8\linewidth]{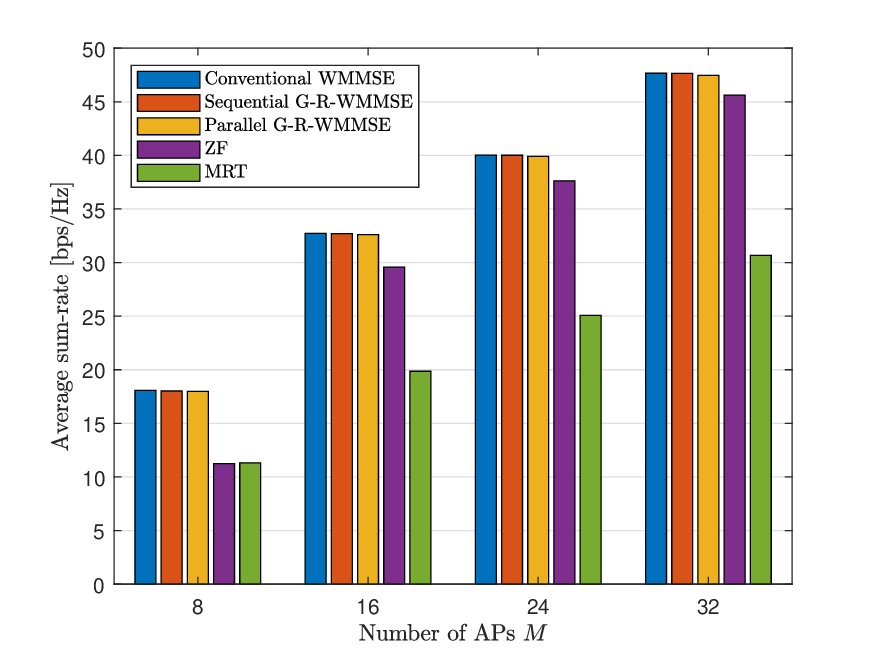}
\vspace{-5mm}
\caption{\label{fig:avgRsum-vs-M}Average sum-rate versus the number of APs $M$ ($K=12$, $n_A=2$, $\text{SNR}^{\text{dl}} = 20$ dB, $\text{SNR}^{\text{ul}} = 10$ dB, and $L=10$).}
\end{figure}

Fig. \ref{fig:avgRsum-vs-M} examines the average sum-rates of the proposed sequential and parallel G-R-WMMSE schemes with respect to the number of APs $M$ with $K=12$, $n_A=2$, $\text{SNR}^{\text{dl}} = 20$, $\text{SNR}^{\text{ul}}  = 10$ dB, and $L=10$.
As benchmarks, we consider the conventional WMMSE algorithm in Sec. \ref{sec:conventional-WMMSE} and closed-form beamforming solutions such as the zero-forcing (ZF) and maximum-ratio transmission (MRT). 
Albeit the lower complexity, the sequential G-R-WMMSE achieves almost identical performance to the conventional WMMSE and outperforms the ZF and MRT methods.
The parallel G-R-WMMSE scheme exhibits a slight performance loss compared to the conventional WMMSE and the sequential G-R-WMMSE schemes.
However, as discussed in Sec. \ref{sec:complexity}, the parallel G-R-WMMSE scheme notably decreases the time complexity.
The performance gap between the ZF scheme and the conventional WMMSE or G-R-WMMSE schemes diminishes as $M$ increases. This reduction in the performance gap is attributed to the expansion of the null space dimension of the interference channel matrix, which is given by $n_A M - K + 1$.

\begin{figure}
\centering\!\!\!\!\!\includegraphics[width=0.8\linewidth]{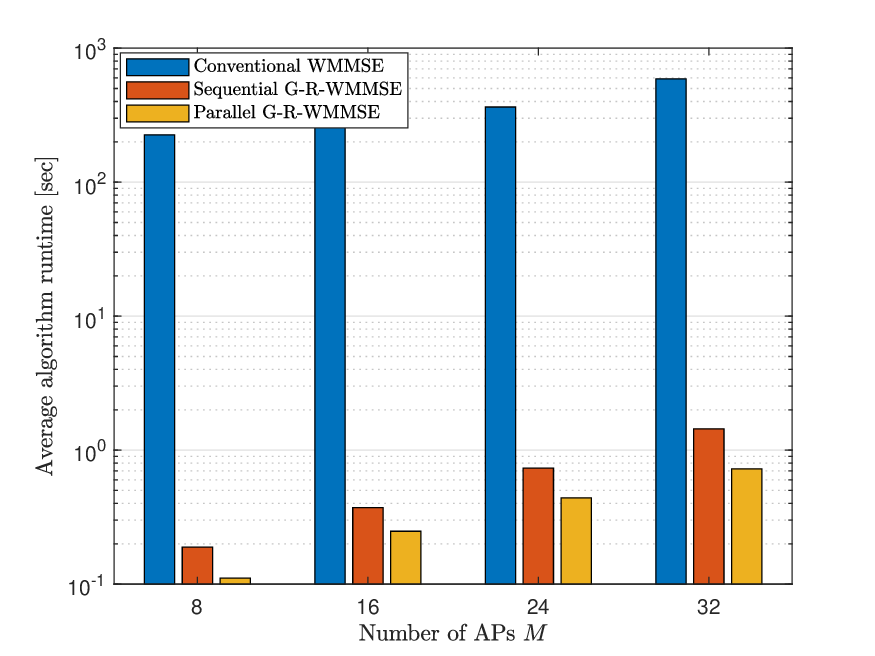}
\vspace{-5mm}
\caption{\label{fig:avgTime-vs-M}Average algorithm runtime versus the number of APs $M$ ($K=12$, $n_A=2$, $\text{SNR}^{\text{dl}} = 20$ dB, $\text{SNR}^{\text{ul}} = 10$ dB, $L=10$).}
\end{figure}

Fig. \ref{fig:avgTime-vs-M} compares the complexity of the conventional WMMSE and the proposed G-R-WMMSE algorithms in terms of the average
algorithm runtime under the same experimental conditions as in Fig. \ref{fig:avgRsum-vs-M}.
The figure illustrates that the proposed G-R-WMMSE scheme achieves complexity savings exceeding 99$\%$, as compared to the conventional WMMSE
scheme. This result is particularly promising noting that the proposed G-R-WMMSE scheme achieves the performance of the conventional WMMSE scheme (as shown in Fig. \ref{fig:avgRsum-vs-M}).

\begin{figure}
\centering\!\!\!\!\!\includegraphics[width=0.8\linewidth]{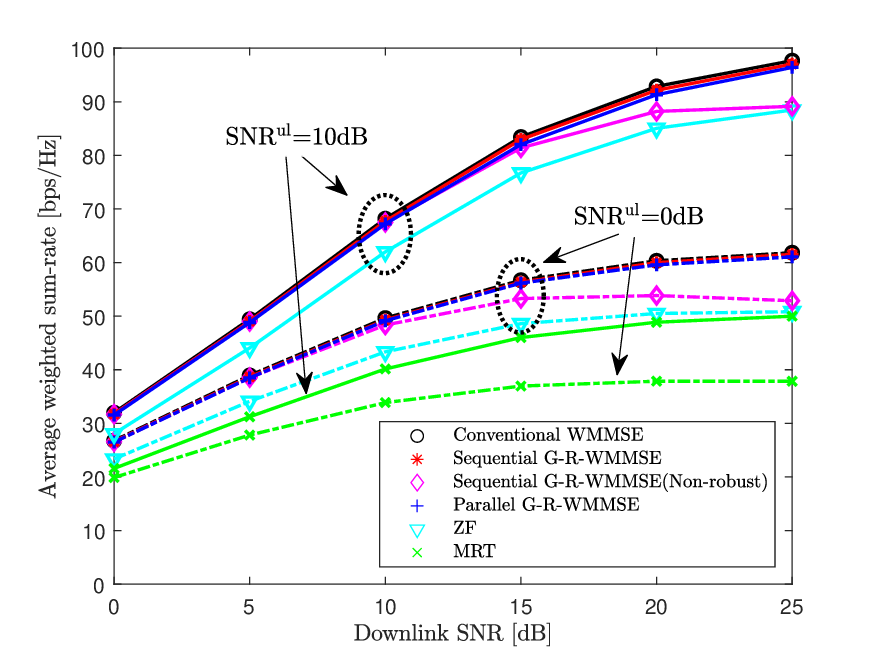}
\vspace{-5mm}
\caption{\label{fig:avgRsum-vs-SNR}Average weighted sum-rate versus the downlink SNR ($K=24$, $M=32$, $n_A=2$, $\text{SNR}^{\text{ul}} \in \{0, 10\}$ dB, and $L=20$).}
\end{figure}

Fig. \ref{fig:avgRsum-vs-SNR} depicts the average weighted sum-rate with respect to the downlink SNR with $K=24$, $M=32$, $n_A=2$, $\text{SNR}^{\text{ul}} \in \{0, 10\}$ dB, and $L=20$.
The weight $\mu_{k}$ is randomly set to 
\begin{align}
    \mu_k = \left(K\tilde{\mu}_k\right)/\left(\sum\nolimits_{l\in\mathcal{K}}\tilde{\mu}_l\right),
\end{align}
where $\tilde{\mu}_k \sim \mathcal{U}(0,1)$ is the uniform random variable.
Similar to the findings in Fig. \ref{fig:avgRsum-vs-M}, the proposed G-R-WMMSE schemes achieve the weighted sum-rate performance of the conventional WMMSE scheme for arbitrary weights. Notably, they consistently outperform the ZF and MRT schemes.
We also present the performance of a non-robust version of the sequential G-R-WMMSE scheme which determines $\mathbf{v}$ assuming that the estimated CSI is perfect, i.e., $\tilde{\rho}_{k,i} = 0$, $\forall k\in\mathcal{K}$, $i\in\mathcal{M}$.
The advantage of the robust design over the non-robust scheme becomes more pronounced with increasing downlink SNR $\text{SNR}^{\text{dl}}$, as the impact of CSI errors becomes less significant compared to the additive noise signals in the low SNR regime of the downlink channel.

\section{Conclusion} \label{sec:conclusion}

We have proposed the G-R-WMMSE scheme for the efficient optimization of cooperative beamforming vectors for cell-free mMIMO systems. 
Beamforming vectors were partitioned across individual APs, which facilitate the BCD approach to optimize the partitioned subvectors jointly.
To tackle the subproblem of optimizing each subvector, 
we derived a closed-form solution by leveraging the Lagrange duality method.
We also presented a parallel implementation of the proposed G-R-WMMSE scheme, which 
allows for parallel computation across the APs and achieves additional complexity savings compared to the sequential scheme.
Simulation results have demonstrated that the proposed G-R-WMMSE schemes achieve the sum-rate performance of the conventional WMMSE scheme with reduced complexity.

\end{document}